\documentstyle[12pt]{article}
\newcommand{\AmS}{{\protect\the\textfont2A\kern-.1667em
\lower.5ex\hbox{M}\kern-.125emS}}

\newcommand {\be} {\begin{eqnarray}} 
\newcommand {\ee} {\end{eqnarray}}

\input{psfig.sty}

\begin{document}

\Huge{\noindent{Istituto\\Nazionale\\Fisica\\Nucleare}}

\vspace{-3.9cm}

\Large{\rightline{Sezione SANIT\`{A}}}
\normalsize{}
\rightline{Istituto Superiore di Sanit\`{a}}
\rightline{Viale Regina Elena 299}
\rightline{I-00161 Roma, Italy}

\vspace{0.65cm}

\rightline{INFN-ISS 96/12}
\rightline{December 1996}

\vspace{1.5cm}

\begin{center} 

\Large{LIGHT-FRONT CQM CALCULATIONS OF\\ BARYON ELECTROMAGNETIC FORM
FACTORS}\footnote{To appear in Nuclear Physics {\bf A} (1997): Proceedings
of the International Workshop on {\em Hadron dynamics with the new DA$\phi$NE
and CEBAF facilities}, Frascati, Italy, 11-14 November 1996.}\\

\vspace{1cm}

\large{F. Cardarelli $^{(1)}$, E. Pace $^{(2)}$, G. Salm\`{e} $^{(3)}$ and
S. Simula $^{(3)}$}

\vspace{1cm}

\normalsize{$^{(1)}$Dept. of Physics and Supercomputer Computations Research
Institute,\\ Florida State University, Tallahassee, FL 32306,USA}

\vspace{0.25cm}

\normalsize{$^{(2)}$Dipartimento di Fisica, Universit\`{a} di Roma "Tor Vergata"
and\\ Istituto Nazionale di Fisica Nucleare, Sezione Tor Vergata,\\ Via della
Ricerca Scientifica 1, I-00133 Roma, Italy}

\vspace{0.25cm}

\normalsize{$^{(3)}$Istituto Nazionale di Fisica Nucleare, Sezione
Sanit\`{a},\\ Viale Regina Elena 299, I-00161 Roma, Italy}

\end{center}

\vspace{1cm}

\begin{abstract}

\noindent The parameter-free predictions for the $N - P_{11}(1440)$
and $N - P_{33}(1232)$ electromagnetic transition form factors, obtained
within our light-front constituent quark model using eigenfunctions of a
baryon mass operator which includes a large amount of configuration
mixing, are reported. The effects due to small components in the baryon
wave functions, such as S'- and D-wave, are also investigated.

\end{abstract}

\vspace{0.5cm}

\newpage

\pagestyle{plain}

\section{INTRODUCTION}

\indent The electromagnetic (e.m.) excitation of Nucleon resonances, in
the space-like region, represents a very interesting tool for gathering
information on the internal structure of baryons and it is one of the
major issues of the TJNAF research programme \cite{TJNAF}. For
investigating this topic we have developed a phenomenological approach
\cite{CAR_N}, based on a constituent quark ($CQ$) model which features: 
i) a proper treatment of relativistic effects, achieved by formulating the
model on the light-front ($LF$), see e.g. \cite{KP91}; ii) baryon
eigenfunctions of a mass operator that describes fairly well the mass
spectrum \cite{CI86}, at variance with the widely adopted gaussian-like
ansatz, see e.g.  Ref. \cite{CK95}; iii) a one-body approximation for the
e.m. current able to reproduce the experimental data on the Nucleon and
pion form factors.

\section{GENERAL FORMALISM}

\indent In the $LF$ hamiltonian dynamics \cite{KP91} a baryon state, in the
$u -d$ sector, $|\Psi_{J J_n ~ \pi}^{T T_3}, ~ \tilde{P} \rangle$, is an
eigenstate of:  i) isospin, $T$ and $T_3$; ii) parity, $\pi$; iii)
kinematical (non-interacting) $LF$ angular momentum operators $j^2$ and
$j_n$, where the vector $\hat{n} = (0,0,1)$ defines the spin quantization
axis; iv) total $LF$ baryon momentum $\tilde{P} \equiv (P^+,
\vec{P}_{\perp}) = \tilde{p}_1 + \tilde{p}_2 + \tilde{p}_3$, where $P^+ =
P^0 + \hat{n} \cdot \vec{P}$ and $\vec{P}_{\perp} \cdot \hat{n}=0$.  The
state $|\Psi_{J J_n ~ \pi}^{T T_3}, ~ \tilde{P} \rangle$ factorizes into
$|\Psi_{J J_n ~ \pi}^{T T_3} \rangle ~ |\tilde{P} \rangle$, and the
intrinsic part $|\Psi_{J J_n ~ \pi}^{T T_3} \rangle$ can be constructed
from the eigenstate $|\psi_{J J_n ~ \pi}^{T T_3} \rangle$ of the {\em
canonical} angular momentum, i.e.  $|\Psi_{J J_n ~ \pi}^{T T_3} \rangle =
{\cal{R}}^{\dag} ~ |\psi_{J J_n ~ \pi}^{T T_3} \rangle$, where the
unitary operator ${\cal{R}}^{\dag} = \prod_{j=1}^3 R_{Mel}^{\dag}
(\vec{k}_{j}, m_j)$, with $R_{Mel} (\vec{k}_{j}, m_j)$ being the
generalized Melosh rotation \cite{KP91}. Then $|\psi_{J J_n ~ \pi}^{T
T_3} \rangle$ satisfies the following mass equation
 \be
    (M_0 + V) ~ |\psi_{J J_n ~ \pi}^{T T_3} \rangle ~ = ~ M ~ |\psi_{J J_n
    ~ \pi}^{T T_3} \rangle
    \label{2}
 \ee
where $M_0 = \sum_{i=1}^3 \sqrt{m_i^2 + \vec{k}_i^2 }$ is the free mass
operator, $m_i$ the $CQ$ mass ($m_u = m_d = 0.220 ~ GeV$ accordingly to
\cite{CI86}), $M$ the baryon mass, and $J(J+1)$, $J_n$ are the eigenvalues
of the operators $j^2$, $j_n$, respectively. The interaction $V$ has to be
independent of the total momentum $P$ and invariant upon spatial rotations
and translations. We can identify Eq. (\ref{2}) with the baryon mass
equation proposed by Capstick and Isgur ($CI$) \cite{CI86}. Their $CQ$
interaction is composed by a linear confining term, dominant at large
separations, and a one-gluon-exchange ($OGE$) term, dominant at short
separations, given by a central Coulomb-like potential and a
spin-dependent part, responsible for the hyperfine splitting of baryon
masses. The mass equation (\ref{2}) has been accurately solved by
expanding the eigenstates onto a large harmonic oscillator ($HO$) basis
(up to $20 ~ HO$ quanta) and then by applying the Rayleigh-Ritz variational
principle.

\indent The $CQ$ momentum distribution calculated from the baryon
eigenfunctions of Eq.  (\ref{2}), with the $CI$ interaction, has a
striking feature \cite{CAR_N}: for a $CQ$ three-momentum larger than
$1~GeV/c$, it is order of magnitude larger than momentum distributions
evaluated from model functions, such as gaussian or power-law fiunctions
(cf. \cite{CK95}). This fact is due to the smoothly singular $OGE$ part
and has immediate consequences on the interpretation of the resonances,
for instance, the Roper resonance is not a simple (first) radial
excitation of the Nucleon and it has a large mixed-symmetry S'component
($P_{S'}^{Roper} = 9.3 \%$).

\indent In the $LF$ formalism the space-like e.m. form factors are
related to the matrix elements of the {\em plus} component of the
current, ${\cal{I}}^+ = {\cal{I}}^0 + \hat{n} \cdot {\vec{\cal{I}}}$,
with the standard choice $q^+ = q^0 + \hat{n} \cdot {\vec{q}} =
P^{+}_f - P^+_i = 0$, that allows to suppress the contribution of the
pair creation from the vacuum \cite{ZGRAPH}.  For a ${1 \over 2}^+$
baryon in the final state, e.g.  the Nucleon ($f=N$) or the Roper
resonance ($f=R$), the Dirac and Pauli form factors, $F_{1(2)}^{f
\tau}(Q^2)$ ($\tau=p$ or $n$), are given by
 \be
    F_1^{f \tau}(Q^2) = {1 \over 2} ~ Tr[{\cal{I}}^+(\tau)] ~~~~ , ~~~~
    F_2^{f \tau}(Q^2) = i {M_f + M_N \over 2Q} ~ Tr[\sigma_2 ~
   {\cal{I}}^+(\tau)]
   \label{4}
 \ee
with ${\cal{I}}^{+}_{\nu_f \nu}(\tau)=\bar{u}_{LF}(\tilde{P}_f, \nu_f)
\left \{ F_1^{f \tau}(Q^2) ~ \gamma^{+} + F_2^{f \tau}(Q^2) ~ i \sigma^{+
\rho} q_{\rho} /( M_f + M_N) \right \} u_{LF}(\tilde{P}_i, \nu)$, where
$Q^2 \equiv - q \cdot q~$ is the squared four-momentum transfer, $\sigma^{+
\rho} = {i \over 2}[\gamma^{+},\gamma^{\rho}]$, $u_{LF}(\tilde{P}_i, \nu)$
[$u_{LF}(\tilde{P}_f, \nu_f)$] the Nucleon [final baryon] LF-spinor,
$\sigma_2$ a Pauli matrix. For the excitation to a $\Delta$ resonance, or
in general to a ${3 \over 2}^+$ baryon, the kinematic-singularity free
form factors $G_{1,2,3}$ \cite{DEK} are related to the LF matrix elements
of ${\cal{I}}^+$ as follows
 \be
    {\cal{I}}^+_{{3 \over 2}{1 \over 2}} & = & {Q \over \sqrt{2}} \left [
    G_1\left(Q^2\right) + {M_{\Delta} - M_N \over 2} G_2\left(Q^2\right )
    \right ] \nonumber \\
    {\cal{I}}^+_{{1 \over 2}{1 \over 2}} & = & -{Q^2 \over \sqrt{6}} \left
    [ {G_1\left(Q^2\right) \over M_{\Delta}} + {G_2\left(Q^2\right) \over
    2}-{M_{\Delta} - M_N \over M_{\Delta}} G_3\left(Q^2\right) \right ]
    \nonumber \\
    {\cal{I}}^+_{{1 \over 2}-{1 \over 2}} & = & {Q \over \sqrt{6}}
    \left [ G_1\left(Q^2\right){M_N\over M_{\Delta}} - {M_{\Delta} - M_N
    \over 2 M_{\Delta}} G_2\left( Q^2\right) -{ Q^2 \over M_{\Delta}}
    G_3\left(Q^2\right) \right ] \nonumber \\
    {\cal{I}}^+_{{3 \over 2}-{1 \over 2}} & = & -{Q^2 \over 2\sqrt{2}}
    G_2\left(Q^2\right)
    \label{6}
 \ee
with ${\cal{I}^+}_{\nu_f \nu}(\tau) = \langle\frac{1}{2}\tau, 10 | TT_3
\rangle ~ \bar{w}^{\mu}_{LF}(\tilde{P}_{f}, ~ \nu_f) ~ \Gamma_{\mu}^{~+} ~
u_{LF}(\tilde{P}_{i}, ~ \nu) $, with $\Gamma_{\mu}^{~+} =
G_1\left(Q^2\right) {\cal{K}}_{1\mu}^{~+}+ G_2\left(Q^2\right)$
${\cal{K}}_{2\mu}^{~+} + G_3\left(Q^2\right){\cal{K}}_{3\mu}^{~+}$
(tensors ${\cal{K}}_{i\mu}^{~\rho}$ are defined in \cite{DEK}).
Differently from the case of a ${1 \over 2}^+$ baryon, the number of form
factors for the excitation of a ${3 \over 2}^+$ baryon is not equal to the
one of the LF matrix elements of ${\cal I}^+$, cf.  Eq. (\ref{6}). For
the exact current the inversion of Eq. (\ref{6}) is unique. Since we adopt
a one-body approximation for ${\cal I}^+$ (see below) different choices of
LF matrix elements can lead to different predictions for the $G_i$ form
factors. In the actual calculation for the N-$\Delta$ transition we
consider two different prescriptions for extracting the $G_i$ form factors
from Eq. (\ref{6}): i) $G_{1}$ and $G_{3}$ are obtained from the first
three equations in (\ref{6}), while $G_{2}$ is directly taken from the
fourth one (prescription I); ii) all the $G_i$ form factors are extracted
from the first three equations (prescription II).

\section{RESULTS}

\indent Elastic and transition form factors have been evaluated using
eigenvectors of Eq. (\ref{2}) and approximating the ${\cal{I}}^+$
component of the e.m. current by the sum of one-body $CQ$ currents
(see \cite{CAR_N}), i.e. 
 \be 
    {\cal{I}}^+(0) \approx \sum_{j=1}^3 ~ I^+_{j}(0) = \sum_{j=1}^3 ~ \left
    ( e_j \gamma^+ f_1^j(Q^2) ~ + ~ i \kappa_j {\sigma^{+ \rho} q_{\rho}
    \over 2 m_j}f_2^j(Q^2) \right )
    \label{5}
 \ee
where $e_j$ ($\kappa_j$) is the charge (anomalous magnetic moment) of the
j-th quark, and $f_{1(2)}^j$ the corresponding Dirac (Pauli) form factor. 
Though the hadron e.m. current has to contain many-body components for
fulfilling both gauge and rotational invariances, we have shown in
\cite{CAR_N}(a) that the $CQ$ form factors can be chosen so that the
effective one-body current (\ref{5}) is able to give a coherent and
accurate description of both the pion and Nucleon experimental form
factors.  After fixing the $CQ$ form factors, we have calculated, {\em
without free parameters}, the N-Roper helicity amplitudes $A_{1 \over
2}^{p(n)}(Q^2)$ and $-S_{1 \over 2}^{p(n)}(Q^2)$, shown in Fig. 1 (see
\cite{CAR_N}(c)). Our results both for $A_{1 \over 2}^{p(n)}(Q^2)$ and
$S_{1 \over 2}^p(Q^2)$ exhibit a remarkable reduction (bringing our
predictions closer to the experimental analyses \cite{PDG,EXP_P11}) with
respect to non-relativistic \cite{HYBRID} as well as relativistic
\cite{CK95} predictions, based on simple gaussian-like wave functions. It
is worth noting that the helicity amplitudes $S_{1 \over 2}^p(Q^2 )$ and
$S_{1 \over 2}^n(Q^2)$ are sizably sensitive to the presence of the
mixed-symmetry $S'$ component in the $CI$ wave function.

\indent The magnetic form factor $G_M^{N-\Delta}\left(Q^2\right)$ and the
ratios $E_1/M_1 = -G_E^{N-\Delta}\left(Q^2\right) / G_M^{N-\Delta}
\left(Q^2\right)$ and $S_1/M_1 = - \left (\sqrt{K^+K^-}/4M^2_{\Delta}
\right)$ $G_C^{N-\Delta}\left(Q^2\right)/G_M^{N-\Delta}\left(Q^2\right)$
(see Ref. \cite{DEK} for the relation with the $G_{i}$ form factors),
calculated within our model {\em without free-parameters} (see also
\cite{CAR_N}(b)) are shown in Fig. 2 (a,b,c), respectively. The
differences between the prescription I and II are not so relevant for
$G_M^{N-\Delta}$ as  they are for $E_1/M_1$ and $S_1/M_1$; however the
effect due to the D-wave component is always small for both prescriptions
(indeed $P_D^{\Delta} = 1.1 \%$).  Although two-body $CQ$ currents are
lacking in the approximation (\ref{5}) it is encouraging that our
effective current can provide an overall description of relatively small
quantities, such as the ratios $E_1/M_1$ ($\approx 5~\%$) and $S_1/M_1$
($\approx 10 ~\%$) for $Q^2$ up to few $GeV/c^2$.  Finally in Fig. 2 (d)
the ratio of $G_M^{N-\Delta}\left(Q^2\right)$ (obtained in prescription
II) and the isovector part of the Nucleon magnetic form factor,
$G^p_M\left(Q^2\right) - G^n_M\left(Q^2\right)$, is shown. It can clearly
be seen that this ratio is largely insensitive to the presence of $CQ$
form factors, whereas it is sharply affected by the spin-dependent part of
the $CI$ potential, which is generated by the chromomagnetic interaction.

In conclusion, we have reported the calculations of the e.m. form
factors for the transition $N - P_{11}(1440)$ and $N - P_{33}(1232)$,
obtained within our approach based on a relativistic $CQ$ model,
baryon eigenfunctions of a mass operator and an effective one-body
$CQ$ current. The results allow an overall description of the data,
but they also indicate the necessity of the introduction of at least a
two-body term in the $CQ$ current, in order to give accurate predictions
for "small" form factors (like, e.g.  $E_1$ or $S_1$ for the N-$\Delta$
transition). Finally, we have shown that the determination of
$G_M^{N-\Delta}\left(Q^2\right)/(G^p_M\left(Q^2\right) -
G^n_M\left(Q^2\right))$ could provide relevant information on SU(6)
breaking effects in N and $\Delta$ wave functions, without substantial
model dependence.

\section{ACKNOWLEDGEMENT} One of the authors (F.C.)  acknowledges a
warm hospitality by Simon Capstick and the partial support by the U.S.
Department of Energy through Contract DE-FG05-86ER40273, and by the
Florida State University Supercomputer Computations Research Institute
(SCRI) which is partially funded by the Department of Energy through
Contract DE-FC05-85ER250000.

\newpage
\begin{figure}[htb]
\psfig{figure=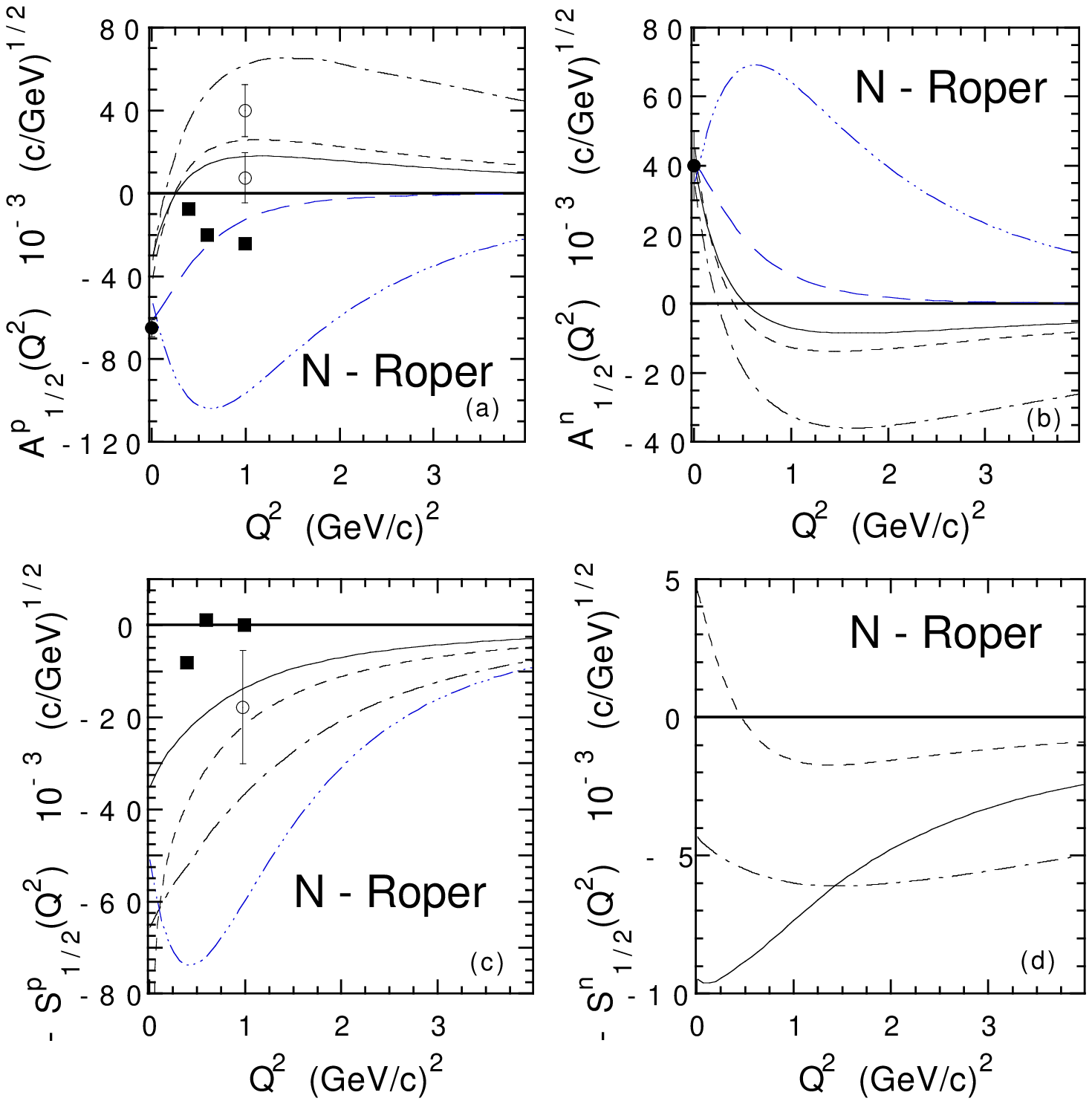,bbllx=-5mm,bblly=140mm,bburx=0mm,bbury=292mm}
Figure 1. {The $N - P_{11}(1440)$ helicity amplitudes $A_{1 \over
2}^{p(n)}(Q^2)$ and $-S_{1 \over 2}^{p(n)}(Q^2)$ vs. $Q^2$. Solid line: 
LF calculation obtained by using baryon wave functions corresponding to
the $CI$ interaction \cite{CI86} and the e.m. current (\ref{5}) with the
$CQ$ form factors determined in \cite{CAR_N}(a); dashed line: the same as
the solid one, but without the S' component in the baryon eigenfunctions;
dot-dashed line: LF calculation obtained by using the gaussian functions
of Ref. \cite{CK95} without CQ form factors. The long-dashed and
triple-dot-dashed lines correspond to the non-relativistic $q^3G$ and
$q^3$ models of Ref. \cite{HYBRID}. Full dots: $PDG$ values \cite{PDG};
full squares and open dots: phenomenological analyses of Ref.
\cite{EXP_P11} (a) and (b), respectively. Within the hybrid $q^3G$ model
$S_{1 \over 2}^{p(n)}(Q^2) = 0$, whereas within the non-relativistic $q^3$
model only $S_{1 \over 2}^n(Q^2)= 0$.  (After \cite{CAR_N} (c)) } 
\end{figure}

\newpage

\begin{figure}[htb]
\psfig{figure=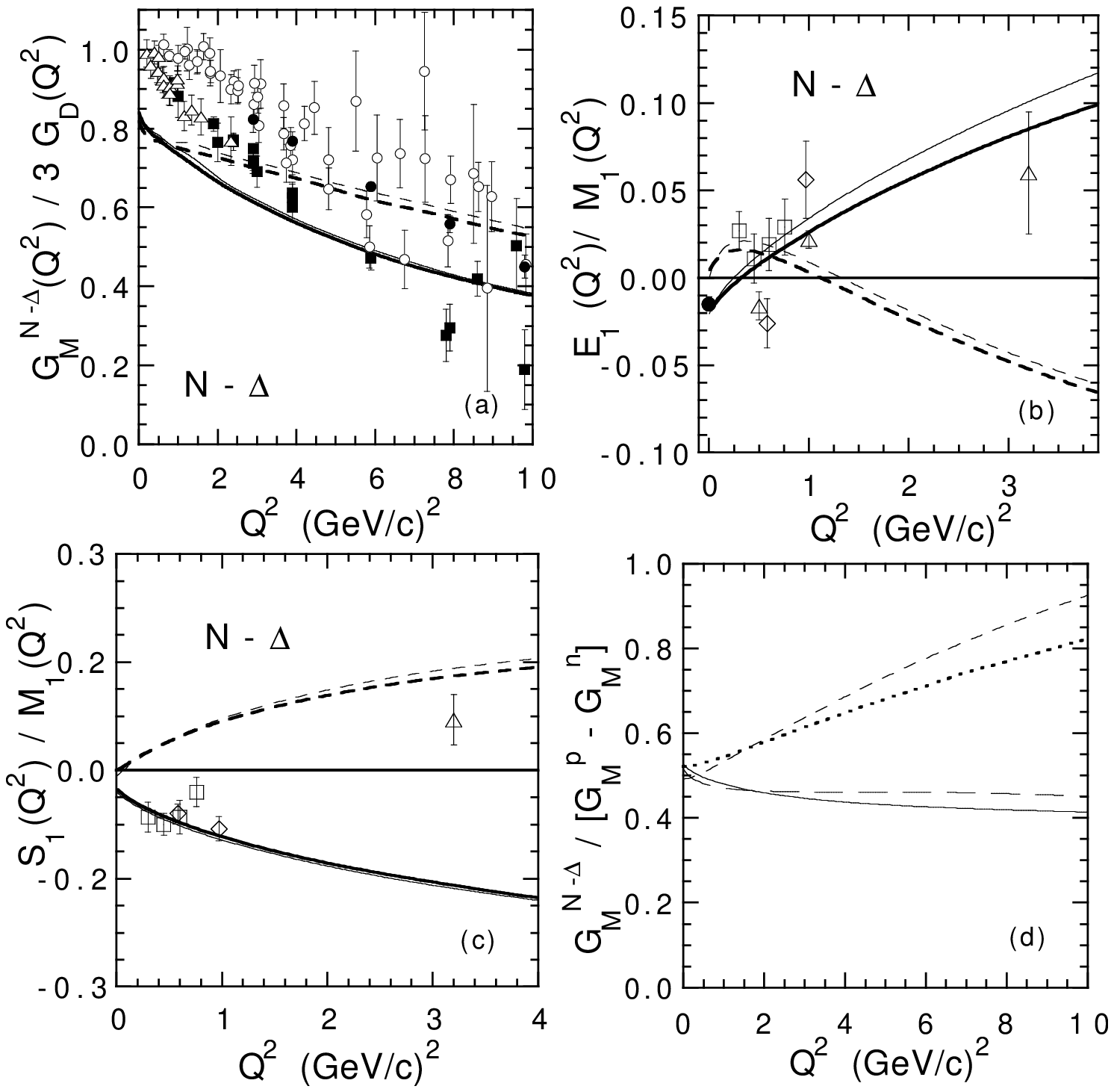,bbllx=-5mm,bblly=140mm,bburx=0mm,bbury=292mm}
Figure 2. {(a) $G_M^{N-\Delta} \left (Q^2 \right ) / 3G_D \left (Q^2
\right )$, with $G_D \left (Q^2 \right ) = 1/(1+Q^2/0.70)^2$, vs $Q^2$. The
thick and thin lines correspond to the LF calculations with and without
the D-wave in the $\Delta$ eigenstate. The e.m. current (\ref{5}) with
the $CQ$ form factors determined in \cite{CAR_N}(a) has been adopted. 
Solid lines: prescription I (see text). Dashed lines: prescription II
(see text). Triangles: \cite{EXP_P33} (a); full squares: \cite{EXP_P33}(b);
open dots:  \cite{EXP_P33}(c); full dots: \cite{EXP_P33}(d). - (b) The
same as in (a), but for $E_1/M_1$. Full dots: PDG \cite{PDG}; diamonds: 
\cite{E1_EXP}(a); open squares: \cite{E1_EXP}(b); triangles: 
\cite{E1_EXP}(c). - (c) The same as in (b) but for $S_1/M_1$. - (d) The
ratio $G_M^{N-\Delta}\left (Q^2 \right )/(G^p_M \left (Q^2 \right )-G^n_M
\left (Q^2 \right ))$ vs $Q^2$. Solid line:  our calculation
(prescription II) with the $CI$ baryon eigenfunctions and $CQ$ form
factors; dashed line: the same as the solid line, but without $CQ$ form
factors; short-dashed line: the same as the dashed line, but with the
baryon eigenfunctions corresponding to the spin-independent part of the
$CI$ interaction \cite{CI86}; dotted line: the same as the dashed line,
but retaining only the confining part of the $CI$ potential. (After
\cite{CAR_N}(b)).}
\end{figure}

\end{document}